\documentclass[aps,amsfonts,amsmath,prd,preprint,nofootinbib]{revtex4}
\newcommand{\beq}{\begin{equation}}
\newcommand{\eeq}{\end{equation}}

\usepackage{csquotes}
\numberwithin{equation}{section}

\usepackage[title]{appendix}

\usepackage{subcaption}

\def\beq{\begin{equation}} 
\def\eeq{\end{equation}}
\def\beqa{\begin{eqnarray}}
\def\eeqa{\end{eqnarray}}
\def\p{\partial}

\usepackage{hyperref} 
\hypersetup{ 
	colorlinks=true, 
	linkcolor=black, 
	filecolor=black, 
	citecolor = blue,       
	urlcolor=blue, 
} 
\usepackage{graphicx}

\begin{document}

\title{Quantum cosmology, eternal inflation, and swampland conjectures}

\author{Georgios Fanaras and Alexander Vilenkin}

\affiliation{Institute of Cosmology, Department of Physics and Astronomy,\\ 
Tufts University, Medford, MA 02155, USA}

\begin{abstract}

In light of the recent swampland conjectures, we explore quantum cosmology and eternal inflation beyond the slow roll regime.  
%the condition under which the universe undergoes stochastic eternal inflation.  
We consider a model of a closed universe with a scalar field $\phi$ in the framework of tunneling approach to quantum cosmology.  The scalar field potential is assumed to have a maximum at $\phi=0$ and can be approximated in its vicinity as $V(\phi)\approx 3H^{2}-\frac{1}{2}m^{2}\phi^{2}$.  Using the instanton method, we find that for $m<2H$ the dominant nucleation channel for the universe is tunneling to a homogeneous, spherical de Sitter space.  For larger values of $m/H$, the most probable tunneling is to an inhomogeneous closed universe with a domain wall wrapped around its equator.  We determine the quantum state of the field $\phi$ in the nucleated universe by solving the Wheeler-DeWitt equation with tunneling boundary conditions.  Our results agree with earlier work which assumed a slow-roll regime $m\ll H$.  We finally show that spherical universes nucleating with $m<2H$ undergo stochastic eternal inflation with inflating regions forming a fractal of dimension $d>2$. For larger values of $m$ the field $\phi$ is unstable with respect to formation of domain walls and cannot be described by a perturbative stochastic approach.

\end{abstract}

\maketitle

\tableofcontents

\section{Introduction}

It is a widely held view that in order to understand how the universe originated we should treat the universe quantum mechanically and describe it by a wave function $\Psi$ rather than by a classical spacetime.  The picture that has emerged from this approach is that a small closed universe can spontaneously nucleate out of nothing, where ‘nothing’ refers to a state with no classical space and time\cite{Vilenkin:1982de,Hartle:1983ai,Linde:1983mx,Rubakov:1984bh,Vilenkin:1984wp,Zeldovich:1984vk}. The cosmological wave function $\Psi$ can be used to calculate the probability distribution for the initial states in the ensemble of nucleating universes.  At least some of these universes are expected to go through a period of inflation, driven by the potential energy $V(\phi)$ of the inflaton field $\phi$. This energy is eventually thermalized, inflation ends, and from then on the universe follows the standard hot cosmological scenario.  Inflation is a necessary ingredient in this kind of scheme, since it provides the only way to get from a tiny nucleated universe to the large universe we live in today.

%Inflationary spacetimes are known to be past-incomplete \cite{?}, indicating that cosmic inflation must have had a beginning.  This raises the question of what determined the initial conditions for inflation.  Even though it may not be essential for observational predictions of inflationary models, this is an important question of principle, and without resolving it the inflationary cosmology remains incomplete.  

%Perhaps the most promising approach to this problem is based on quantum cosmology, which suggests that a spatially closed universe can spontaneously nucleate out of `nothing', where `nothing' refers to a state with no classical space, time and matter \cite{AV82,HH,Linde84,Rubakov84,AV84,ZS}.  The initial conditions for inflation are determined by 

The wave function of the universe $\Psi$ satisfies the Wheeler-DeWitt (WDW) equation
\beq
{\cal H}\Psi = 0,
\label{WDW}
\eeq
where ${\cal H}$ is the Hamiltonian operator.  To solve this equation, one has to choose some boundary conditions for $\Psi$.  The most developed proposals for this choice are the Hartle-Hawking \cite{Hartle:1983ai} and the tunneling \cite{Vilenkin:1986cy,Vilenkin:2018dch} boundary conditions.  The predictions resulting from the two proposals have mostly been studied assuming 
%spherical geometry of the universe and an inflaton field $\phi$ with 
a slowly varying potential $V(\phi)$,\footnote{ We assume for simplicity that the potential depends on a single scalar field $\phi$.  An extension to multiple fields is briefly discussed in Section 2.} 
\beq
|V'|/V\ll 1,~~~|V''|/V\ll 1
\label{slow}
\eeq
in reduced Planck units $(8\pi G =1)$. 
%and that $0\leq V(\phi)\ll 1$.  
%The universe is then predicted to nucleate with a nearly homogeneous field $\phi$ and a probability distribution
%\beq
%P(\phi)\propto \exp(\pm ...)
%\eeq
%for its initial values.  Here the upper and lower signs are for the Hartle-Hawking and tunneling wave functions, respectively.  
Then the Hartle-Hawking wave function predicts that the universe is most likely to nucleate with the lowest (positive) value of $V(\phi)$ and thus disfavors inflation.  The tunneling wave function, on the other hand, predicts nucleation with $\phi$ near the maximum of $V(\phi)$, which is a favorable initial condition for (hilltop) inflation.  Here we shall adopt the tunneling proposal.  

With this choice, and assuming that the conditions (\ref{slow}) are satisfied, most of the universes nucleate with $\phi$ close to the maximum of $V(\phi)$ and immediately enter the regime of stochastic eternal inflation \cite{Vilenkin:1983xq,Linde:1986fd,Starobinsky:1986fx}.  The dynamics of the field $\phi$ near the top of the potential is dominated by quantum fluctuations and is well approximated by a random walk.  In any horizon region the field gradually drifts away from the top and eventually rolls down to the minimum of $V(\phi)$, but in different regions this happens at different times, and at any time there are parts of the universe that are still inflating with $\phi$ near the top.  The inflating regions form a self-similar fractal \cite{Aryal:1987vn}, with their total volume growing exponentially with time. 

This scenario, however, is now being seriously questioned.  There is growing evidence that a wide class of quantum field theories do not admit a UV completion within the theory of quantum gravity, even though they look perfectly consistent otherwise.  Such theories are said to belong to the swampland, as opposed to the landscape.  A number of different criteria that the landscape potential $V(\phi)$ should satisfy have recently been suggested \cite{Obied:2018sgi,Agrawal:2018own}.  One of them is the so-called refined swampland conjecture \cite{Garg:2018reu,Ooguri:2018wrx}, which asserts that at any value of $\phi$ with $V(\phi)>0$ one of the two conditions has to be satisfied:
\beq
|V'|\geq cV,~~~{\rm or} ~~~V''\leq -{\tilde c} V,
\label{swampland}
\eeq
where $c$ and ${\tilde c}$ are positive constants of order one.  This conjecture is clearly incompatible with the conditions (\ref{slow}).  It is also in considerable tension with the inflationary scenario, which usually requires a slowly varying potential $V(\phi)$.

The swampland conjecture (\ref{swampland}) as it stands may or may not be true.  We take an agnostic attitude to this issue, but the above discussion 
%Inflationary cosmology has impressive observational support, and at this time there seem to be no 
%viable alternatives.  It seems reasonable therefore to assume that the swampland criteria, whatever 
%their final form will be, must be consistent with slow-roll inflation.  It is possible for example that the 
%constants $c,c'$ in Eq.~(\ref{swampland}) have somewhat smaller values, e.g., $\sim 0.1-0.01$ \cite{??}.  This 
%would be compatible with slow-roll inflation, even though the possible form of the inflaton potential 
%would be strongly restricted.  Depending on the values of $c$ and ${\tilde c}$, quantum random walk of the 
%inflaton field and the associated eternal inflation may or may not be excluded \cite{??}.  Some authors 
%have even suggested that 
%regardless of the validity of any specific form of the swampland conjecture, 
%eternal inflation may be forbidden by some fundamental principle \cite{Dvali:2018fqu,Dvali:2018jhn,Rudelius:2019cfh}.
%Since the swampland conjecture (\ref{swampland}) may be true with some values of $c$ and ${\tilde c}$ not too different from ${\cal O}(1)$, this 
motivates us to reconsider quantum cosmology and eternal inflation in the regime where the slow variation conditions (\ref{slow}) are not satisfied.  This is our goal in the present paper.  We start with quantum cosmology in Section 2.  The tunneling wave function is expected to be peaked near the maximum of the potential, where $V(\phi)$ can be approximated as
\beq
V(\phi)\approx V_0 -\frac{1}{2}m^2\phi^2.
\label{V}
\eeq
We have $|V''|/V(0)=m^2/3H^2$, where $H=(V_0/3)^{1/2}$ is the expansion rate of a de Sitter universe of vacuum energy density $V_0$.  The slow variation condition (\ref{slow}) is then $m\ll H$.  Here we are mostly interested in the regime $m\gtrsim H$.

We start in Section 2 by discussing nucleation of the universe using the instanton method.  
For sufficiently small values of the mass parameter, $\mu\equiv m/H<2$, we find that nucleation is described by the homogeneous de Sitter instanton having the geometry of a 4-sphere.  The initial state is then a spherical universe with $\phi=0$, as usually assumed.  On the other hand, for $\mu>2$ the dominant instanton describes nucleation of a "domain wall universe."  In this case the initial state still has the topology of a 3-sphere, but is inhomogeneous, having a domain wall centered on a 2-sphere (the `equator') where $\phi=0$.  
We briefly discuss possible scenarios of subsequent evolution of a domain wall universe, referring to appropriate literature for more detail.     

Focusing on the spherical initial configuration, we then discuss 
%For $0\leq\mu\leq 2$ the most likely initial configuration is a spherical universe with $\phi\approx 0$. 
the quantum state of the nucleated universe in Sec.3.  We expand the scalar field $\phi({\bf x},t)$ in spherical modes on a 3-sphere and solve the WDW equation for the wave function $\Psi(a,\phi_n)$, where $a$ is the scale factor and $\phi_n$ are the mode amplitudes.  Of special interest is the homogeneous mode $n=1$.  We find that the probability distribution for $\phi_1$ is fully consistent with the instanton analysis.
For small values of $\mu$ the wave function is peaked at $\phi_1=0$, while for $\mu>2$ this behavior is modified and the top of the potential becomes a local minimum of $|\Psi|$.  We argue that this marks a transition to a domain wall universe.  For inhomogeneous modes, $n>1$, the quantum state is an analytic continuation of the standard Bunch-Davies vacuum to a tachyonic field (with a possible exception of the dipole mode $n=2$).

%We argue that in this regime the most probable initial state is a `domain wall universe'.  It still has the topology of a 3-sphere, but is inhomogeneous, having a domain wall centered on a 2-sphere (the `equator') where $\phi=0$.  We note that $m>2H$ is the condition for domain wall solutions to exist in a de Sitter space of expansion rate $H$ \cite{Basu:1992ue}.  We also discuss possible scenarios of subsequent evolution of a domain wall universe.     
We finally discuss under what conditions nucleation of the universe is followed by stochastic inflation.
One expects that there is some value $\mu_{max}$ above which stochastic inflation is not possible.  
%Eternal character of inflation has a dramatic effect on the structure of spacetime.  It is therefore important to determine the precise conditions under which inflation is eternal.  
Several attempts have been made to determine this threshold value in the literature with different results \cite{Rudelius:2019cfh,Kinney:2018kew,Barenboim:2016mmw}, so the question is still open at this time.  We address this issue in Sec.4, with the conclusion that $\mu_{max}=2$ -- that is, stochastic inflation occurs for all values of $\mu$ that allow nucleation of a spherical universe.  

Our conclusions are summarized and discussed in Sec.5.  
%In particular, we briefly discuss extension of our analysis to models with multiple scalar fields 
%and discuss the relation of our results to previous work.

\section{Instanton analysis}

We consider a model of a scalar field $\phi$ minimally coupled to gravity.  
%In reduced Planck units $8\pi G=1$ 
The action of the model is
\beq
S=\int \sqrt{-g}d^{4}x\left(\frac{R}{2}-\frac{1}{2}(\nabla\phi)^{2}-V(\phi) \right).
\eeq
We assume that the potential $V(\phi)$ has a maximum at $\phi=0$, where it can be approximated as in Eq.(\ref{V}):
\beq
V(\phi)=3H^{2}\left(1-\frac{1}{6}\mu^{2}\phi^{2}\right),
\label{Vmu}
\eeq
where $\mu=m/H$.

In the leading order of semiclassical approximation, nucleation of the universe can be described using the instanton method.  According to the tunneling prescription, the nucleation probability is then given by
\beq
P_{nuc}\propto e^{- |S_E|},
\eeq
where $S_E$ is the action of the Euclidean instanton solution. 
%and the upper and lower signs are for the tunneling and Hartle-Hawking wave functions respectively.  Here we adopt the tunneling prescription.
The dominant instanton is the one with the smallest magnitude of the action $|S_E|$.  It is usually assumed \cite{Vilenkin:1982de,Hartle:1983ai,Linde:1983mx,Rubakov:1984bh,Vilenkin:1984wp} that this is the homogeneous de Sitter instanton with $\phi=0$ and the geometry of a 4-sphere of radius $H^{-1}$.  The corresponding Euclidean action is
\beq
S_E^{(dS)}=-\frac{8\pi^2}{H^2}.
\eeq
We will now argue that this assumption is valid only for sufficiently small values of the mass parameter: $\mu<2$.  For larger values of $\mu$ the dominant instanton is an inhomogeneous domain wall instanton.

Domain wall instantons were first discussed in Refs.\cite{Basu:1991ig,Basu:1992ue} in the context of domain wall nucleation in the inflationary universe. They generally have $O(4)\times Z_2$ symmetry and the topology of a 4-sphere.
Here we will mostly be interested in the regime of $\mu \approx 2$, corresponding to the transition between de Sitter and domain wall instantons.  In this case the instanton solution covers only a small range of $\phi$ near $\phi=0$, so we can use the approximate form of the potential (\ref{Vmu}) and the geometry is accurately described by the Euclideanized de Sitter metric
\beq
ds^2=H^{-2}\{d\zeta^2+\cos^2\zeta[d\chi^2+\sin^2\chi(d\theta^2+\sin^2\theta d\phi^2)]\}.
\eeq
The domain wall instanton 
%has $O(4)$ symmetry and 
can be found with the ansatz $\phi=\phi(\zeta)$.  The field equation for $\phi$ is then
\beq
\phi''-3\tan\zeta \phi'+\mu^2\phi=0
\eeq
with boundary conditions $\phi'(-\pi/2)=\phi'(\pi/2)=0$.  With $\mu\approx 2$, this equation has two solutions: $\phi=0$ and
\beq
\phi(\zeta)\approx A\sin\zeta
\label{phizeta}
\eeq
with $A={\rm const}$.  The first solution corresponds to de Sitter instanton and the second to domain wall instanton.

The constant $A$ depends on the nonlinear terms in the potential $V(\phi)$ \cite{Basu:1992ue}. Its particular value will not be important for our discussion, but as an illustrative example let us include in the potential (\ref{Vmu}) an additional quartic term:
\beq  
V(\phi)=3H^2-\frac{1}{2}H^2\mu^2\phi^2+\lambda\phi^4.
\eeq
Substituting this and the solution (\ref{phizeta}) into the Euclidean action
\beq
S_E=2\pi^2 H^{-4} \int_0^\pi d\zeta \cos^3\zeta \left[\frac{1}{2}(\phi')^2+V(\phi)\right],
\eeq
we find
\beq
S_E=\frac{2\pi^2}{H^2}\left[3H^2-\frac{2}{15}H^2(\mu^2-4)A^2+\frac{4}{35}\lambda A^4\right].
\eeq
This action is minimized by 
\beq
A^2=\frac{7}{12\lambda}(\mu^2-4)
\label{A}
\eeq
for $\mu>2$ and by $A=0$ for $\mu<2$.  Note that the domain wall instanton solution (\ref{phizeta}), (\ref{A}) does not exist for $\mu<2$.  Note also that this threshold value depends only on $\mu$ and is not sensitive to the nonlinear part of the potential.
A numerical calculation in Ref.\cite{Basu:1992ue} confirms that $|S_E^{(wall)}| <|S_E^{(dS)}|$ for all $\mu>2$.\footnote{ This is similar to Coleman-de Luccia instanton \cite{CdL} describing bubble nucleation in false vacuum.  This instanton solution is dominant when it exists, but it disappears and is replaced by the Hawking-Moss instanton \cite{HM} when the tunneling barrier becomes sufficiently flat near the top \cite{Hiscock}.}

Negative modes of the de Sitter instanton have been discussed in the Appendix of Ref.\cite{Basu:1992ue}.  It shows that a homogeneous negative mode $\phi={\rm const}$ is always present for $\mu^2>0$, as it should be in a tunneling instanton.  There are no other negative modes for $\mu<2$, but for $\mu>2$ additional, dipole negative modes of the form (\ref{phizeta}) appear.  They signal that the de Sitter instanton is no longer dominant and that the leading instanton has a domain wall structure.  

\begin{figure}[t]
  \begin{subfigure}[t]{0.475\textwidth}
    \includegraphics[scale=0.4]{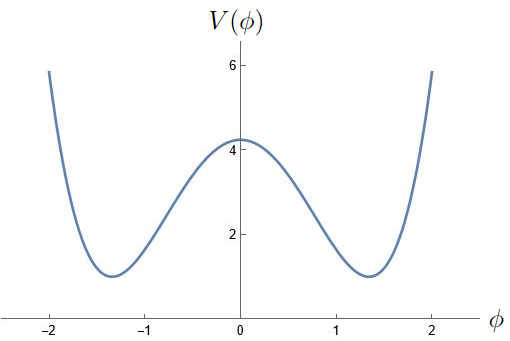}
    \caption{Two de-Sitter vacua symmetric about $\phi=0$.}
    \label{fig-1a}
  \end{subfigure}\hfill
  \begin{subfigure}[t]{0.475\textwidth}
    \includegraphics[scale=0.4]{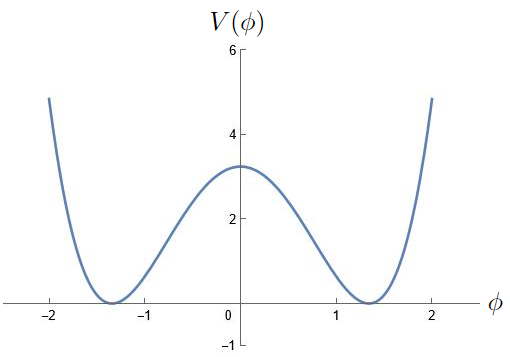}
    \caption{Two Minkowski vacua symmetric about $\phi=0$.}
    \label{fig-1b}
  \end{subfigure}
  \begin{subfigure}[t]{0.475\textwidth}
    \includegraphics[scale=0.4]{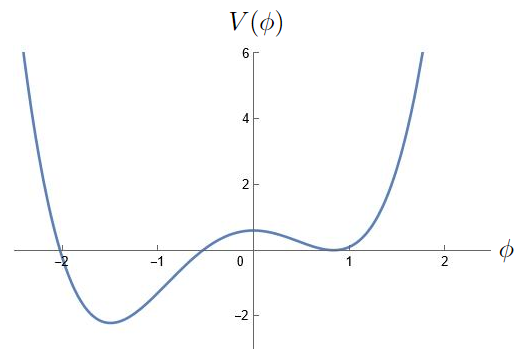}
    \caption{An anti de-Sitter vacuum at $\phi<0$ and a Minkowski vacuum at $\phi>0$.}
    \label{fig-1c}
  \end{subfigure}
  \caption{A plot of $V(\phi)$ with $\mu>2$ and a local maximum at $\phi=0$ in between two local minima.} 
  \label{fig1}
\end{figure}

We conclude that the dominant nucleation channel in models with $\mu>2$ is to a domain wall universe.  
Possible scenarios for subsequent evolution of such a universe depend on the shape of the potential $V(\phi)$ away from $\phi=0$; they have been discussed in Ref.\cite{Blanco-Pillado:2019tdf}.  A few characteristic examples are illustrated in Fig.\ref{fig1}.

In Fig.\ref{fig-1a} the potential is a double well with a maximum at $\phi=0$ sandwiched between two symmetric de Sitter minima.  Then at late times we have an approximately spherical expanding de Sitter universe with a domain wall on its equator.  An important point is that domain walls of a fixed physical thickness exist in de Sitter space only for $\mu>2$ \cite{Basu:1993rf, Bonjour:1999kz}.  In the opposite case, wall-like configurations like Eq.(\ref{phizeta}) are smeared by the expansion of the universe, evolving towards a homogeneous $\phi=0$ solution.  In Fig.\ref{fig-1b} the minima of the potential are at $V=0$.  The universe then evolves to an inflating domain wall solution with two asymptotically Minkowski regions bounded by an expanding domain wall \cite{Ipser:1983db}.  More precisely, the regions on the two sides of the wall have the geometry of open $(k=-1)$ FRW universes.
The minima of $V(\phi)$ do not have to be symmetric; an example is shown in Fig.\ref{fig-1c}.  Once again, the wall separates two regions containing open FRW universes evolving with $\phi>0$ and $\phi<0$ branches of the potential.  With a suitable choice of parameters, the potential can have a flat portion which can lead to a long inflationary period; an example is discussed in \cite{Blanco-Pillado:2019tdf}.  The potential minimum at $\phi<0$ in Fig.\ref{fig-1c} is at $V<0$.  As a result the FRW universe on that side of the wall develops a big crunch singularity \cite{Abbott}.

We finally briefly discuss multi-field models where the potential can be approximated near its maxima as
\beq
V(\phi)=3H^{2}-\frac{1}{2}\sum_{i=1}^{N}m_{i}^{2}\phi_{i}^{2}.
\eeq
Instanton solutions for such models in the de Sitter background have been discussed in \cite{Basu:1992ue} assuming that the potential has an $O(N)$ symmetry, so that all mass parameters $m_i^2$ are equal to one another.  For $N=2$ the de Sitter instanton dominates if $m_i<2H$.  Otherwise, the dominant nucleation channel is to a `global string universe', having a global string wrapped around its equator.   Similarly, for $N=3$ with $m_i>2H$, the dominant tunneling is to a `global monopole universe', having a global monopole and anti-monopole at its opposite poles.  For models with lower symmetry the nucleated universe would have a more complicated defect structure.  For example, in an $N=2$ model with $Z_2\times Z_2$ symmetry we would have $m_1\neq m_2$ and the global string would be attached to four walls.

\section{Quantum state of the universe}

\subsection{General formalism}

%The cosmological wavefunction with tunneling boundary conditions has been defined in the slow-roll regime $\mu\ll1$ \cite{Vilenkin:1986cy}. Here we will extend this analysis, but we will keep $\mu\phi\ll1$ such that the expansion about the maximum of the potential is justified.

We shall first review the standard minisuperspace analysis of the tunneling wave function.  It was originally performed in the slow-roll regime $\mu\ll1$ (e.g., \cite{Vilenkin:1987kf,Vachaspati:1988as}). Here we will extend this analysis, but we will keep $\mu^2\phi^2\ll1$, so that the expansion about the maximum of the potential is justified.

The metric is assumed to be homogeneous, isotropic and compact:
\beq
ds^{2}=a(\eta)^{2}\left(-d\eta^{2}+d\Omega_{3}^{2}\right)
\eeq
where $a(\eta)$ is the scale factor, $\eta$ is the conformal time and $d\Omega_{3}$ is the metric on a unit 3-sphere. 

We expand our field $\phi$ in spherical harmonics $Q$:
\beq
\phi=(2\pi^{2})^{1/2}\sum_{n,l,m}\phi_{nlm}(t)Q^{n}_{lm}(\textbf{x}),
\eeq
where $(2\pi^{2})^{1/2}$ is a normalization factor and $n,~l,~m$ are harmonic numbers on a 3-sphere. For further convenience we will denote all harmonic numbers as $n$. 

The Wheeler-DeWitt (WDW) equation for the model is obtained through the quantization of the Hamiltonian constraint. In this case it recasts to
\beq
\left[\frac{1}{24\pi^{2}}\frac{\p^{2}}{\p a^{2}}-6\pi^{2}V(a)-\sum_{n}\mathcal{H}_{n}\right]\Psi(a,\phi_{n})=0,
\label{wdw}
\eeq
where we have neglected the factor ordering ambiguity and
\beq
V(a)=a^{2}\left(1-H^{2}a^{2}\right),
\eeq
\beq
\mathcal{H}_{n}=\frac{1}{2a^{2}}\frac{\p^{2}}{\p\phi_{n}^{2}}-\frac{1}{2a^{2}}\omega_{n}^{2}\phi_{n}^{2},
\eeq
\beq
\omega_{n}^{2}=(n^{2}-1)a^{4}-m^{2}a^{6},
\eeq
where $n\in\{1,2,3...\}$. 

The next step is to solve the WDW equation. We will treat the scalar field modes $\phi_{n}$ as small perturbations. Then a solution to (\ref{wdw}) can be obtained with the ansantz
\beq
\Psi\propto \exp\left[-12\pi^{2}S_{0}(a)-\frac{1}{2}R_{n}(a)\phi_{n}^{2}\right].
\label{PsiWDW}
\eeq
Substituting the above to (\ref{wdw}) and neglecting subdominant terms in the WKB expansion along with terms of order $\mathcal{O}\left(\phi_{n}^{4}\right)$, we have
\beq
\left(\frac{dS_{0}}{da}\right)^{2}-V(a)=0
\label{eqS0}
\eeq
and
\beq
a^{2}\left(\frac{dS_{0}}{da}\right)\left(\frac{dR_{n}}{da}\right)-R_{n}^{2}+\omega_{n}^{2}=0.
\label{eqRn}
\eeq
The solution to (\ref{eqS0}) is straightforward and given by
\beq
\frac{dS^{\pm}_{0}}{da}=\pm\sqrt{V(a)}\longrightarrow S_{0}^{\pm}(a)=\mp\frac{\left(1-H^{2}a^{2}\right)^{3/2}}{3H^{2}}
\eeq
where $S_{0}^{\pm}(a)$ correspond respectively to the decreasing and growing branches in the under-barrier region $Ha<1$. In the classically allowed region $Ha>1$ they are the contracting and expanding branches.

We now turn our attention to (\ref{eqRn}). We make a change of variables
\beq
\frac{da}{d\tau}=
    \begin{cases}
      \sqrt{V(a)} & \text{$\tau<0$}\\
      -\sqrt{V(a)} & \text{$\tau>0$}
    \end{cases}  
\eeq
where $\tau$ is the Euclidean conformal time related to $\eta$ through $\eta=i\tau$ and
\beq
a=\frac{1}{H\cosh(\tau)}.
\eeq
With this change (\ref{eqRn}) becomes 
\beq
a^{2}\frac{dR^{\pm}_{n}}{d\tau}-{R^{\pm}_{n}}^{2}+\omega_{n}^{2}=0.
\eeq
This is a Riccati equation and can be linearized via the transformation
\beq
R^{\pm}_{n}(\tau)=-\frac{a^{2}}{\nu_{n}}\frac{d\nu_{n}}{d\tau}.
\eeq
The resulting ODE recasts to
\beq
\ddot{\nu}_{n}+2\frac{\dot{a}}{a}\dot{\nu}_{n}-\left(n^{2}-1-m^{2}a^{2}\right)\nu_{n}=0,
\label{modeeq}
\eeq
where we substituted the explicit form of $\omega_{n}$.

The solutions to the above can be expressed in terms of the associated Legendre functions:
\beq
\nu_{n}(\tau)=A_{n}\cosh\tau~P^{n}_{\gamma+1}(\tanh\tau)+B_{n}\cosh\tau~Q^{n}_{\gamma+1}(\tanh\tau),
\label{modes}
\eeq
where
\beq
\gamma=\left(\frac{9}{4}+\frac{m^2}{H^2}\right)^{1/2}-\frac{3}{2}
\eeq
and $A_{n},~B_{n}$ are constant coefficients.
 %that will have to be determined by the boundary conditions. 
 The functions $\nu_{n}$ correspond to the mode functions of a tachyonic field $\hat{\phi}$ of mass $im$ in the de-Sitter background geometry. Selecting the integration constants in (\ref{modes}) is equivalent to selecting the quantum state of the scalar field $\hat{\phi}$.

The tunneling boundary conditions yield a total of three branches: growing and decaying under the barrier and expanding in the allowed region. The under the barrier components $\Psi_{+}$ and $\Psi_{-}$ are to be evaluated for $\tau<0$ and $\tau>0$ respectively. The expanding branch in the Lorentzian region can be found by analytically continuing $\Psi_{-}$ to imaginary conformal time $\eta=i\tau$.
%the allowed region through the conformal time $\eta$. Note, however, that the 
%specification of the coefficients 
The coefficients multiplying the exponentials in Eq.(\ref{PsiWDW}) can then be determined by matching at the turning point $\tau=\eta=0$ \cite{Vachaspati:1988as}.

\subsection{Regularity conditions}

In order to specify the tunneling wave function we will need to determine the arbitrary coefficients $A_{n},~B_{n}$ in Eq.(\ref{modes}).  This can be done by requiring that the wave functions for the mode amplitudes $\phi_n$ in (\ref{PsiWDW}) for all three branches are Gaussian rather than inverse Gaussian: 
\beq
R^{\pm}_{n}(a)\geq 0~,~~{Re} \{R_{n}(a)\}\geq 0.
\label{regularity}
\eeq
Here, $R^{+}_{n}$ and $R^{-}_{n}$ are for the decaying and growing branches respectively and $R_{n}$ is for the expanding branch.  (Note that $R_n^\pm$ are real.)  Following earlier literature, we will call Eq.(\ref{regularity}) the regularity condition.  It ensures that the wave function is peaked at $\phi_n=0$, so that field fluctuations are suppressed.

%Since we are dealing with a tachyonic field the appropriate boundary conditions should ensure regularity of the 3 branches such that the scalar field fluctuations are suppressed. This is the expected behavior of the tunneling prescription in which the wavefunction favors the maxima of the scalar potential $V(\phi)$.

%The regularity condition \textbf{for the three branches} is explicitly expressed as:
%\beq
%R^{\pm}_{n}(a)\geq0~,~~R_{n}(a)\geq0
%\eeq
%where the $R^{\pm}_{n}$ is the decaying and growing branches respectively and $R_{n}$ is the expanding branch. These requirements must hold at least in the slow-roll regime $\mu\ll1$. The tunneling boundary conditions for the case of a tachyonic scalar field have been spelled out in \cite{Wang:2019spw} . Here we will follow a similar procedure. 

We begin by examining the behavior of the growing branch well inside the Euclidean region, $Ha\ll 1$.\footnote{ Here we follow the analysis in Ref.\cite{Wang:2019spw}.}  
In the limit $\tau\rightarrow+\infty$ the mode functions $\nu_{n}$ behave as\footnote{This expansion along with additional properties of Legendre functions which will be used throughout the paper can be found in \href{https://dlmf.nist.gov/}{DLMF} or obtained with Mathematica.}
\beq
\nu_{n}(\tau)\approx \tilde{A}_{n}~e^{(1-n)\tau} + \tilde{B}_{n}~e^{(n+1)\tau}
\eeq
where we have absorbed numerical factors in $\tilde{A}_{n}$ and $\tilde{B}_{n}$. For $B_n\neq 0$, the dominant term is the second one, so that $\nu_{n}\propto e^{(n+1)\tau}$. This would give
\beq
R^{-}_{n}(a)=-a^{2}\frac{d\ln\nu_{n}}{d\tau}\approx -(n+1)a^{2}<0.
\eeq
The regularity condition can be satisfied if we set $B_n=0$; then
\beq
R^-_n(a)\approx (n-1)a^2 \geq 0.
\eeq
This suggests that we set
\beq
\nu_{n}(\tau)=\cosh\tau~P^{n}_{\gamma+1}(\tanh\tau).
\label{modesBD}
\eeq

We next check regularity at the turning point $(a=H^{-1},~\tau=0)$.  Continuity of the wave function at this point requires that
\beq
\lim_{\tau\rightarrow0^{+}}R^{-}_{n}(\tau)=\lim_{\tau\rightarrow0^{-}}R^{+}_{n}(\tau)=\lim_{\eta\rightarrow0^{+}}R_{n}(\eta)
\eeq
The Legendre functions and their derivatives are continuous at the origin. This implies that the constants $A_{n}$ and $B_{n}$ ought to be identical and the values of $R_n(\tau=0)$ should be the same for all three branches.  

It is shown in Ref.\cite{Wang:2019spw} that with the mode functions (\ref{modesBD}) the regularity condition at the turning point is satisfied for all inhomogeneous modes $(n>1)$ with $\mu<2$.  Furthermore, it is shown in Ref.\cite{Damour:2019iyi} that the regularity condition is satisfied in the entire classically allowed range $Ha>1$ if it is satisfied at the turning point.
However, the homogeneous mode $n=1$ has $R_1(a=H^{-1})<0$ for all $\mu^2 > 0$.  At smaller values of $a$ it also develops a singularity on the dominant branch: $R_1^+(a\to a_1)\to -\infty$ at some point $a_1$ in the range $0<a_1<H^{-1}$.  This mode therefore requires a special treatment.

To find the mode function suitable for the homogeneous mode, we shall use the results of Refs.\cite{Vilenkin:1986cy,Vilenkin:1987kf} for slow-roll potentials as a guide.  They suggest that in the regime $Ha\ll1$ and $\mu\phi_1\ll1$, the relevant part of the decaying branch should behave as
\beq
\Psi_{+}(Ha\ll1)\propto\exp\left[-6\pi^{2}a^{2}-\frac{\pi^{2}H^{2}a^{4}}{4}\mu^{2}\phi_1^{2}\right],
\label{Ha>>1mf<<1}
\eeq
{where we kept the leading terms in $a$ and $\phi_1$.}  Comparing this with (\ref{PsiWDW}) we have
%In our analysis we can make the following matching:
\beq
S^{+}_{0}(Ha\ll1)\approx \frac{a^{2}}{2}~~,~~~R^{+}_{1}(Ha\ll1)\approx \frac{\pi^{2}H^{2}a^{4}}{2}\mu^{2},
\label{Rslowroll}
\eeq
where the second expression is valid for $\mu\ll1$. 
%\textbf{(I had made a mistake previously and took the limit $Ha\gg1$. I corrected the expressions. It does not affect the following analysis. GF)}

On the other hand, the homogeneous mode function at $\tau\rightarrow-\infty$ (or $Ha\ll1$) will generally behave as
\beq
\nu_{1}(\tau\rightarrow-\infty)\propto \left(2A_{1}\sin\gamma\pi+B_{1}\pi\cos\gamma\pi \right)~e^{-2\tau}=C_{1}e^{-2\tau},
\label{nu1tau}
\eeq
where we defined
\beq
C_{1}=2A_{1}\sin\gamma\pi+B_{1}\pi\cos\gamma\pi .
\eeq
For $C_{1}\neq0$ this gives
\beq
R^{+}_{1}(Ha\ll1)=-a^{2}\frac{d\ln\nu_{1}}{d\tau}\approx 2a^{2},
\eeq
which disagrees with (\ref{Rslowroll}).  An agreement is achieved only if we set
\beq
B_{1}=-A_{1}\frac{2}{\pi}\tan(\gamma\pi),
\eeq
so that $C_1=0$.  
%It is verified in \cite{Wang:2019spw} that with this choice the homogeneous mode satisfies the regularity condition for all $\mu^2>0$.

%Even though the regularity condition is satisfied at $Ha\ll1$ it will be violated near the turning point. Furthermore, our calculation makes no assumption about the range of values of $\mu$ so we would expect it to agree with expression (\ref{Ha>>1mf<<1}). This can only be achieved if the leading order terms cancel out:
%\beq
%C_{1}=0\longrightarrow B_{1}=-A_{1}\frac{2}{\pi}\tan(\gamma\pi)
%\eeq

\begin{figure}[h!]
		\includegraphics[scale=0.5]{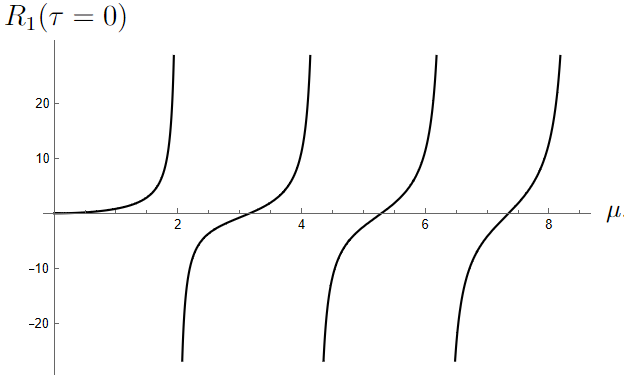}
		\caption{A plot of $R_{1}(\tau=0)$ with respect to $\mu$. For $\mu<2$ regularity holds. As larger values of $\mu$ it still holds for some subregions.}
		\label{fig2}
 
\end{figure}

\begin{figure}[h!]
		\includegraphics[scale=0.5]{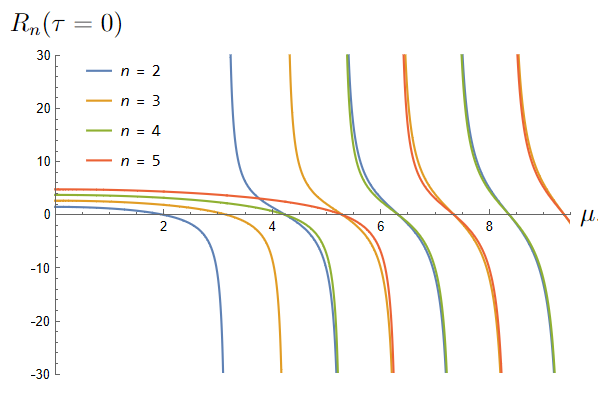}
		\caption{A plot of $R_{n}(\tau=0)$ vs. $\mu$ for the modes $n={2,3,4,5}$. Higher order modes violate regularity at larger values of $\mu$. This leads to a pattern of interference in the pockets of regularity after $\mu>2$.}
		\label{fig3}
 
\end{figure}

The value of $R_1^+(0)$ for the above choice of the homogeneous mode is plotted as a function of $\mu$ in Fig.\ref{fig2}.  We see that regularity is satisfied for $\mu<2$ and violated in some range above $\mu=2$.  This violation of regularity first occurs precisely at $\mu=2$, when the de Sitter instanton becomes subdominant.
At still larger values of $\mu$, $R_1^+(0)$ becomes positive in some subregions.  However, inhomogeneous modes with $n>1$ follow regularity patterns that interfere with the pockets of regularity of the homogeneous mode.  This becomes clear from Fig.\ref{fig3} where we plot $R_{n}(0)$ for a number of leading inhomogeneous modes.  When the mode $n=1$ becomes regular at $\mu>2$, higher order modes violate regularity.
%The existence of these pockets of regularity leaves open the possibility that the wave function may be regular for arbitrarily large $\mu$. However, before we reach such a conclusion it is necessary to investigate the behavior of inhomogeneous modes with $n>1$. 
%We follow the same procedure as with the homogeneous mode and evaluate $R_{n}(\tau=0)$. 

So far we have only verified the regularity condition at $a\to 0$ and $a\geq H^{-1}$.  By sampling different values of $\mu$ numerically, we convinced ourselves that regularity holds at $\mu<2$ on both growing and decaying branches in the entire classically forbidden range $0\leq a\leq H^{-1}$.  There are however two exceptions to this rule.

The first exception is the homogeneous mode $n=1$ on the growing branch, which has  $R_1^-(a) <0$ in a range of $a$ adjacent to $a=0$ for all $\mu^2>0$.  
%This is not surprising: the homogeneous mode on this branch behaves as the Hartle-Hawking wave function, which has a minimum at the maximum of $V(\phi)$.  
The other exception is the dipole mode $n=2$, which violates regularity on the decaying branch for $1.9\lesssim \mu<2$. These regularity violations are discussed in Appendix A, where we argue that they are not necessarily dangerous.

%We thus conclude that regularity is satisfied for all $\mu<2$ and violated for $\mu>2$, where the de Sitter instanton is subdominant.  

Overall, our analysis suggests the following choice of the mode functions:\beq
\nu_{n}(\tau)=
    \begin{cases}
      \cosh\tau~\left[P^{1}_{\gamma+1}(\tanh\tau)-\frac{2}{\pi}\tan(\gamma\pi)~Q^{1}_{\gamma+1}(\tanh\tau)\right] & \text{$n=1$}\\
      \cosh\tau~P^{n}_{\gamma+1}(\tanh\tau) & \text{$n>1$}
    \end{cases}  
    \label{tunnelmodes}
\eeq
We note that inhomogeneous modes follow the usual Bunch-Davies prescription, the only difference being that we are now dealing with a tachyonic field. As a result the modes with $n>1$ will be regular even if we analytically continue to a field with a positive mass $(\mu^2<0)$. The same does not hold for the homogeneous mode which would yield
\beq
R^{+}_{1}(Ha\ll1)\approx -\frac{\pi^{2}H^{2}a^{4}}{2}|\mu^{2}|.
\eeq
%divergent branches once analytically continued
%\footnote{It was suggested in \cite{Wang:2019spw} that one can use the Bunch-Davies $n=1$ mode for the case of a massive scalar field. We note that this prescription is problematic since it will lead a discontinuity of the mode functions after analytic continuation of the mass parameter.}. 
This is a clear characteristic of the tunneling wave function, for which the probability density is maximized at the maxima and minimized at the minima of the scalar potential.\footnote{Note that this conclusion differs from that in Ref.\cite{Garriga:1996gr} which discussed a free massive scalar field $(\mu^2<0)$ model using the Bunch-Davies mode function for the homogeneous mode.  This mode function is regular for a non-tachyonic field discussed in that paper, but would exhibit pathological behavior in more general models that would include tachyonic fields at the maxima of the potential.}

On the other hand, the appropriate choice for the Hartle-Hawking state is to select the Bunch-Davies modes (\ref{modesBD}) for all $n$.
%\footnote{\textbf{This is true for $m^2 > 0$.  Otherwise, the HH wave function is not "regular": it has a minimum at the maximum of $V(\phi)$.}}
This gives a regular wave function for $\mu^2<0$ -- that is, for an ordinary (non-tachyonic) massive field.
%For a tachyonic field with $\mu^2>0$, the homogeneous mode violates regularity -- as it should, since the Hartle-Hawking wave function is peaked at the minima and minimized at the maxima of the potential.
Note that the difference between the tunneling and Hartle-Hawking approaches is limited only to the behavior of the homogeneous mode, and thus they predict the same quantum state for scalar field perturbations.

\section{Condition for eternal inflation}

Suppose now that the universe nucleates with $\phi$ near the top of the potential, where $V(\phi)$ is well approximated by Eq.(\ref{Vmu}) with $\mu<2$, and has initial geometry close to de Sitter.  We want to determine the conditions under which the universe will then undergo stochastic eternal inflation.

\subsection{From Fokker-Planck equation}

We shall first review the case of a small mass, $\mu\ll 1$, which has been extensively studied in the literature.  The scalar field dynamics in this case is accurately described by the Fokker-Planck equation \cite{Starobinsky:1986fx}
\beq
\frac{\p\rho}{\p t}=\frac{H^3}{8\pi^2}\frac{\p^2 \rho}{\p\phi^2}+\frac{1}{3H}\frac{\p}{\p\phi}\left(\rho \frac{\p V}{\p\phi}\right).
\label{FP}
\eeq
The field $\phi$ here should be understood as its average over a horizon volume $\sim H^{-3}$, and the quantity $\rho(\phi,t)d\phi$ has the meaning of the probability for $\phi$ to be in the range $[\phi, \phi+d\phi]$ at time $t$.  The first term on the right-hand side of (\ref{FP}) accounts for quantum diffusion of the field and the second term for the classical drift due to the potential.  Comparing the two terms and using $V(\phi)$ from Eq.(\ref{Vmu}) and `naive' estimates like $\p V/\p\phi\sim V/\phi$, we find that quantum fluctuations dominate at 
\beq
|\phi|\ll\phi_q\equiv H^2/m.
\label{range}
\eeq
The eternally inflating region (EIR) of spacetime can be defined as the region where $\phi$ is in this range.
We note that for such values of $\phi$
\beq
\frac{m^2\phi^2}{V_0}\ll H^2 \ll 1,
\label{H<1}
\eeq
where the last inequality assumes that the potential energy density at the top of the potential is small compared to Planck density.  It follows from Eq.(\ref{H<1}) that the geometry of EIR is well approximated by de Sitter space of energy density $V_0$. 

Outside of EIR scalar field dynamics is nearly classical, so $\phi$ rolls down from the top of the potential and the geometry significantly deviates from de Sitter.  However, this exterior region can affect the geometry of EIR only within a small distance $\sim H^{-1}$ from its boundary.  The 3-volume of EIR at a late time $t$ can be estimated as
%probability for a randomly picked comoving point in the nucleated universe to be in the EIR at time $t$ is
\beq
{\cal V}_{EIR}(t) \sim {\cal V}_0 \exp(3Ht) \int_{-\kappa\phi_q}^{\kappa\phi_q} \rho(\phi,t) d\phi ,
\label{VEIR}
\eeq
where $\kappa\gg 1$, ${\cal V}_0=2\pi^2 H^{-3}$ is the initial volume at nucleation and $\exp(3Ht)$ is the volume expansion factor of inflating regions.  The parameter $\kappa$ depends on our definition of EIR; its precise value is not important for our discussion here, while the condition $\kappa\gg1$ ensures that  the integration range in (\ref{VEIR}) covers the region where quantum fluctuations of $\phi$ are non-negligible.

With initial condition $\rho(\phi,0)=\delta(\phi)$, the solution of the Fokker-Planck equation is given by a Gaussian distribution
\beq
\rho(\phi,t)=(2\pi)^{-1/2}\sigma^{-1}(t)\exp[-\phi^2/2\sigma^2(t)].
\label{Gaussian}
\eeq
with \cite{Aryal:1987vn}
\beq
\sigma^2(t) = \frac{3H^2}{8\pi^2 \mu^2}\left[\exp\left(\frac{2}{3}\mu^2 Ht\right)-1\right].
\label{sigma}
\eeq
In the limit of large $t$ it approaches a $\phi$-independent form
\beq
\rho={\rm const}\cdot \exp(-\mu^2 Ht/3)  
\eeq
and Eq.(\ref{VEIR}) gives
\beq
{\cal V}_{EIR}(t) \propto \exp(dHt),
\label{VEIRt}
\eeq
where
\beq
d=3-\mu^2/3.
\eeq
It was shown in Ref.\cite{Aryal:1987vn} that the eternally inflating region in this case is a self-similar fractal of fractal dimension $d$.

The solution (\ref{Gaussian}), (\ref{sigma}) was used by Rudelius \cite{Rudelius:2019cfh} to determine the condition for eternal inflation (see also references in \cite{Rudelius:2019cfh} for earlier attempts in this direction).  Requiring that the inflating volume grows with time, he found from Eq.(\ref{VEIRt}) the condition $\mu<3$.  We note however that the Fokker-Planck equation on which the above solution is based is applicable only for slowly varying potentials satisfying the conditions (\ref{slow}), which in our case correspond to $\mu\ll 1$.   Here we need to explore the regime of $\mu\sim 1$, so we cannot rely on the Fokker-Planck approximation.

\subsection{From quantum cosmology}

Our approach will be based on quantum cosmology.  The wave function of the universe in the classically allowed region $Ha>1$ is given by Eq.(\ref{PsiWDW}) 
\beq
\Psi\propto \exp\left[-12\pi^{2}S_{0}(a)-\frac{1}{2}R_{n}(a)\phi_{n}^{2}\right]
\label{PsiWDW1}
\eeq
with
\beq
R_{n}(\eta)=-\frac{a^{2}}{\nu_{n}}\frac{d\nu_{n}}{d\eta},
\eeq
where $a(\eta)=-1/H\eta$ and the mode functions $\nu_n(\eta)$ are obtained by analytic continuation $\tau\to i\eta$ from Eq.(\ref{modes}).  In this section it will be more convenient to use the proper time $t=-H^{-1}\ln(-H\eta)$ instead of the conformal time $\eta$.

The mode amplitudes $\phi_n$ behave as independent Gaussian variables; hence the probability distribution for $\phi_n$ can be written as
\beq
dP(\{\phi_n\},t)=\prod_n d\phi_n \rho_n(\phi_n),
\eeq
where
\beq
\rho_n(\phi_n,t)=(2\pi)^{-1/2}\sigma_n^{-1}(t)\exp[-\phi_n^2/2\sigma_n^2(t)].
%\label{Gaussian}
\eeq
From Eq.(\ref{PsiWDW1}) it is clear that the variances $\sigma_n(t)$ are given by
\beq
\sigma_n^2(t)=\frac{1}{2Re(R_{n}(t))}
\label{vari}
\eeq
The mode variances $\sigma_n$ are calculated in Appendix B.

The total field $\phi$, comprising all the modes, depends linearly on the mode amplitudes; hence its probability distribution is also Gaussian:
\beq
dP(\phi,t)=\rho(\phi,t)d\phi=(2\pi)^{-1/2}\sigma^{-1}(t)\exp[-\phi^2/2\sigma^2(t)] d\phi,
\label{Gaussian1}
\eeq
where
\beq
\sigma^2=\langle\phi^2\rangle=\sum_n n^2 \sigma_n^2 
\label{sigmatot}
\eeq
and $n^2$ is the mode degeneracy factor.  

As it stands, the sum in Eq.(\ref{sigmatot}) is divergent at large $n$.  This is not surprising: quantum fluctuations of $\phi$ on scale $l$ grow as $l$ is decreased, so a calculation of the expectation value of $\phi^2$ at a point always gives a divergent result.   This divergence can be regulated by smoothing the field operator ${\hat\phi}({\bf x}, t)$ over horizon-size regions $(l\sim H^{-1})$.  We can do this using a Gaussian window function, or more conveniently by cutting off the summation over $n$ in (\ref{sigmatot}) at some $n_* \sim e^{Ht}$.

The probability distribution for a smoothed field is still given by the Gaussian distribution (\ref{Gaussian1}).
To analyze the regime of eternal inflation, we will be interested in this distribution in the limit of $t\to\infty$.
We found in {Appendix B} that in this limit the time dependence of $\sigma_n$ is the same for all $n$:
\beq
\sigma_n^2\propto \exp(2\gamma Ht),
\eeq
where
\beq
\gamma=\left(\frac{9}{4}+\mu^2\right)^{1/2}-\frac{3}{2},
\eeq
as before.  Hence
\beq
\sigma^2=\sum_{n=1}^{n_*} n^2 \sigma_n^2 =CH^{2}\exp(2\gamma Ht).
\label{sigmaC}
\eeq
The dimensionless constant $C$ is estimated in Appendix B:
\beq
C\approx \frac{2^{1+2\gamma}}{\sin(\gamma\pi)}\frac{\left[\Gamma\left(\gamma+\frac{3}{2}\right)\right]^{2}}{\Gamma\left(\gamma+1\right)\Gamma\left(\gamma+3\right)}.
\eeq
In principle, $C$ should depend on the cutoff $n_{*}$. However, we show in Appendix B that large-$n$ terms in Eq.(\ref{sigmatot}) make a negligible contribution to the series, making $C$ effectively independent of $n_*$.  This indicates that $\sigma^2$ is not sensitive to the size of the smoothing region. We note that $C$ is divergent at $\gamma=1$ $(\mu =2)$ when the modes become unstable. Another divergence is observed in the massless limit $\gamma\rightarrow0$ $(\mu\to 0)$ which corresponds
to the infrared singularity in de-Sitter space (see \cite{Linde:1982uu,Starobinsky:1982ee,Vilenkin:1982wt}).

We can now estimate the volume of EIR in the same way as we did in the preceding subsection:
\beq
{\cal V}_{EIR}\sim {\cal V}_0 \int_{-\kappa\phi_q}^{\kappa\phi_q} \rho(\phi,t) d\phi={\cal V}_0 \frac{2\kappa\phi_q}{H\sqrt{C}}e^{-\gamma Ht}.
\eeq
The result depends on the definition of EIR (parameter $\kappa$), but the fractal dimension $d$ of the inflating region depends only on the time dependence of the inflating volume:
\beq
{\cal V}_{EIR}\propto e^{(3-\gamma)Ht}\equiv e^{dHt}.
\eeq
This gives
%In the limit of large $t$ 
%$\sigma^2(t) \propto \exp(2\gamma Ht)$, where\textbf{, again:}
%\beq
%\gamma=\left(\frac{9}{4}+\frac{m^2}{H^2}\right)^{1/2}-\frac{3}{2}.
%\eeq
%Then the volume of the inflating region is given by Eq.(\ref{VEIRt}) and its fractal dimension is
\beq
d=3-\gamma=\frac{9}{2}-\left(\frac{9}{4}+\mu^2\right)^{1/2}.
\label{d}
\eeq
It follows that for $\mu < 2$ we have stochastic inflation with fractal dimension $d>2$.  As $\mu$ approaches $2$ from below, the fractal dimension approaches that of a domain wall, $d=2$.  And as we discussed in Sec.2, for $\mu>2$ the dominant instanton describes nucleation of a domain wall universe.

We derived the fractal dimension (\ref{d}) assuming that the universe originated by quantum nucleation from nothing.  This result, however, is more general; it is not sensitive to the assumption about the initial state.
For example, Guth and Pi \cite{Guth:1985ya} calculated the scalar field variance in de Sitter space with flat spatial sections assuming a thermal initial state.  They found the same asymptotic time dependence of $\langle\phi^2\rangle$ as in (\ref{sigmaC}), and this would give the same fractal dimension of EIR as in (\ref{d}).\footnote{ We note that $\mu<2$ was also suggested as the condition for topological inflation in which inflation takes place inside the cores of domain walls \cite{Linde:1994hy,Vilenkin:1994pv}. For numerical simulations of the fractal structure of such models see \cite{Linde:1994wt}.} 
This conclusion would formally also apply for $\mu>2$, suggesting $d<2$.  However, in this case de Sitter space admits stable domain wall solutions, so fluctuations of the field $\phi$ are unstable with respect to formation of domain walls.
These are nonlinear structures; they inflate along the wall surfaces and keep a fixed thickness in the transverse direction.  Thus a description in terms of stochastic inflation is appropriate only for $\mu<2$.

\section{Conclusions}

In this paper we studied quantum cosmology and eternal inflation in the regime where the slow-roll conditions are violated.  
Our approach was based on the wave function of the universe with tunneling boundary conditions.  
We considered a model of a closed universe with a scalar field $\phi$, focusing on the range of $\phi$ near the top of its potential, where the potential can be approximated as
$V(\phi)\approx 3H^2 - (1/2)m^2\phi^2$.  We were mostly interested in the regime $m\gtrsim H$.  Our analysis shows that the physics of both nucleation of the universe and eternal inflation undergo a sharp transition at the critical value of $m=2H$.  

We first used the instanton method in Section 2 to study nucleation of the universe from nothing.  For small values of the steepness parameter $\mu=m/H$ we find that the dominant nucleation channel is to a spherical de Sitter universe with the scalar field at the top of the potential, $\phi=0$. As the value of $\mu$ increases and reaches the threshold $\mu=2$, another possible nucleation channel appears -- to an inhomogeneous domain wall universe having a domain wall wrapped around its equator.  This channel, when it exists, always dominates the nucleation probability.  These conclusions are largely based on the work \cite{Basu:1992ue} in the context of domain wall nucleation in de Sitter space (which is described by the same instanton as nucleation of a domain wall universe).  This earlier work, however, did not make a connection to quantum cosmology.  The subsequent evolution of a domain wall universe depends on the nature of the vacua separated by the wall (which are generally different).  We discussed possible scenarios for the cases of (A)dS and Minkowski vacua, following earlier discussion in Ref.\cite{Blanco-Pillado:2019tdf}.  
%With a suitable potential $V(\phi)$, these scenarios may include a period of inflation consistent with observations.

The quantum state of a nucleated spherical universe is discussed in Section 3, where we extend earlier treatments which assumed a slow-roll potential.  To this end, we expand the field in spherical harmonics and solve the Wheeler-DeWitt equation for the wave function of the universe $\Psi(a,\phi_n)$ with tunneling boundary conditions.  
%(Here $a$ is the scale factor and $\phi_n$ are the mode amplitudes.)
The tunneling prescription instructs us to impose the regularity condition, ensuring that $|\Psi|$ decreases as the mode amplitudes $\phi_n$ are increased.  We find a regular quantum state for $\mu<2$, while for larger values of $\mu$ such states do not exist.  These results are consistent with the instanton approach. 

A caveat in the above analysis is the behavior of the dipole mode which violates regularity in the narrow range $2>\mu\gtrsim1.9$.  This feature is discussed in Appendix A, where we argue that it is not necessarily problematic.  We note that the origin of this behavior remains enigmatic for us and it should, perhaps, be further investigated. 

Eternal inflation in the nucleated universe is discussed in Section 4.  The standard treatment using the
Fokker-Planck equation is not applicable beyond the slow-roll regime.  Our approach here is based on quantum cosmology and does not suffer from this limitation. The wave function of the universe provides 
Gaussian probability distributions for individual modes $\phi_{n}$.  We use this to find the distribution for the total field $\phi$ (comprising all the modes) smoothed over horizon regions, and then use it in turn to estimate the volume of the eternally inflating region (EIR).  We find that for $\mu<2$ this volume grows exponentially with time, indicating that eternal inflation does occur in this range of $\mu$.  The rate of volume growth is related to the fractal dimension $d$ of EIR, and we find that $d>2$ and that $d\to 2$ from above as $\mu\to 2$ from below.  For $\mu>2$ stable domain wall solutions exist in de Sitter space, and fluctuations of the field $\phi$ are unstable with respect to formation of walls.  A description in terms of stochastic inflation is not appropriate in this regime.

As a final remark, we comment on the status of inflation in relation to swampland conjectures.  Inflationary cosmology has impressive observational support and at this time there seem to be no 
viable alternatives.  It seems reasonable therefore to assume that the swampland criteria, whatever 
their final form will be, must be consistent with slow-roll inflation.  It is possible for example that the 
constants $c,c'$ in Eq.~(\ref{swampland}) have somewhat smaller values, e.g., $\sim 0.1-0.01$ \cite{Kinney:2018nny}.  This would be compatible with slow-roll inflation, even though the possible form of the inflaton potential would be strongly restricted.  Depending on the values of $c$ and ${\tilde c}$, quantum random walk of the inflaton field and the associated eternal inflation may or may not be excluded \cite{Matsui:2018bsy,Brahma:2019iyy,Wang:2019eym}.  Some authors have even suggested that regardless of the validity of any specific form of the swampland conjecture, 
eternal inflation may be forbidden by some fundamental principle \cite{Dvali:2018fqu,Dvali:2018jhn,Rudelius:2019cfh}. As stated in the introduction we take an agnostic approach towards these views. However, let's assume for a moment that stochastic eternal inflation is forbidden. The most probable scenario is then a domain wall universe whose walls will inflate indefinitely spawning new inflationary regions \cite{Blanco-Pillado:2019tdf}. 
%Regardless in which vacuum the field will eventually roll into, this process will go on forever and an infinite amount of walls will be formed. This strongly suggests that eternal inflation is unavoidable in the current cosmological picture, despite the restrictions of the swampland conjectures.
It would be interesting to explore the issues related to anthropic selection and to the measure problem in this new kind of multiverse.

\begin{acknowledgements}
We are grateful to Gia Dvali for a stimulating discussion and to Jose Blanco-Pillado and Jaume Garriga for useful comments on the initial draft of the paper.  This work was supported in part by the National Science Foundation under grant No. 2110466.
\end{acknowledgements}

\begin{appendices}

\section{Anomalous behavior of $n=1,2$ modes}

\begin{figure}[h!]
		\includegraphics[scale=0.43]{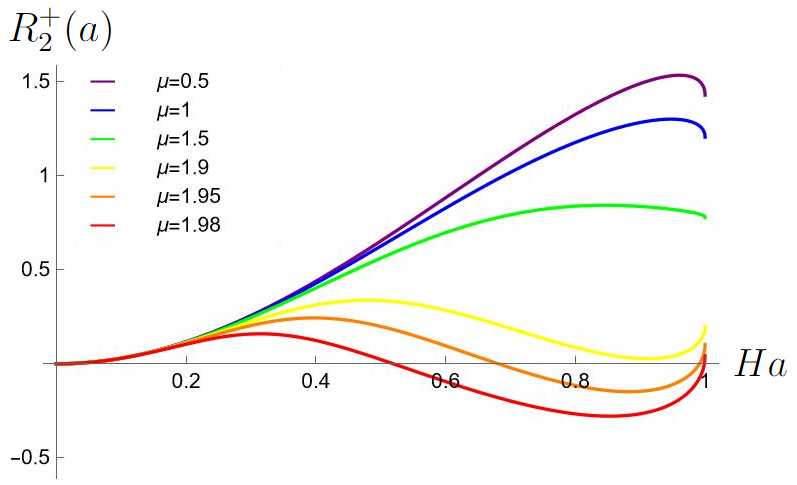}
		\caption{A plot of $R_{2}^+(a)$ for several values of $\mu$. Lower curves correspond to higher values of $\mu$, with the yellow curve which nearly violates regularity corresponding to $\mu=1.9$.}
	
				\label{fig4}
\end{figure}

We noted at the end of Sec. 3.2 that the homogeneous mode $n=1$ violates regularity on the growing branch and the dipole mode $n=2$ violates regularity on the decaying branch.
The latter violation occurs in an intermediate range of $a$ for $1.9\lesssim\mu<2$ (see Fig.\ref{fig4}).  We will argue that these violations are not necessarily dangerous, after discussing a related issue below.

Our selection of the mode functions in Eq.(\ref{tunnelmodes}) was partly based on the analysis of the growing branch of the wave function $\Psi_-(a,\phi_n)$.  However, this branch is exponentially suppressed compared to the decreasing branch (at $\phi_n=0$), so keeping it while we neglect larger corrections to the WKB formula requires some justification.  This issue was addressed in Refs.\cite{Vachaspati:1988as,Wang:2019spw} with the following argument. First note that the growing and decreasing branches have comparable magnitudes near the turning point, $Ha \sim 1$,  so keeping the growing branch is justified in this range. Furthermore, it was argued in \cite{Wang:2019spw} that 
\beq
P_{n}(a)\equiv \frac{R^{-}_{n}(a)}{R^{+}_{n}(a)}<1
\label{P<1}
\eeq
in the entire classically forbidden region $Ha<1$.  This means that the function $\Psi_+(a,\phi_n)$
 decreases with $\phi_n$ exponentially faster than $\Psi_-(a,\phi_n)$, and therefore $\Psi_-$ dominates
at sufficiently large $\phi_n$. We thus have a continuous domain in the $\{a,\phi_n\}$ superspace, ranging
from $a=H^{-1}$ to $a = 0$, where $\Psi_-$ is non-negligible compared to $\Psi_+$.  The inclusion of the growing branch can therefore be extended all the way to $a=0$.
The condition (\ref{P<1}) was verified near the turning point in Ref.\cite{Wang:2019spw}.  Here we used numerical methods to extend this analysis to the entire range $Ha<1$.

\begin{figure}[t]
  \begin{subfigure}[t]{0.475\textwidth}
    \includegraphics[scale=0.4]{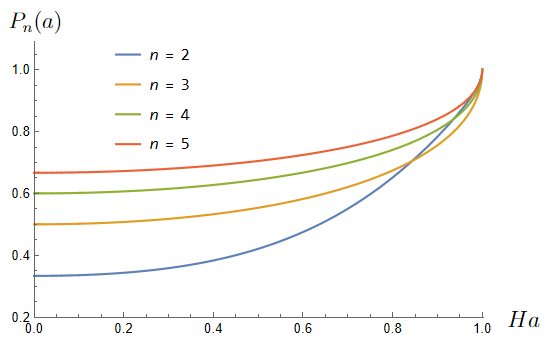}
    \caption{}
    \label{fig-5a}
  \end{subfigure}\hfill
  \begin{subfigure}[t]{0.475\textwidth}
    \includegraphics[scale=0.41]{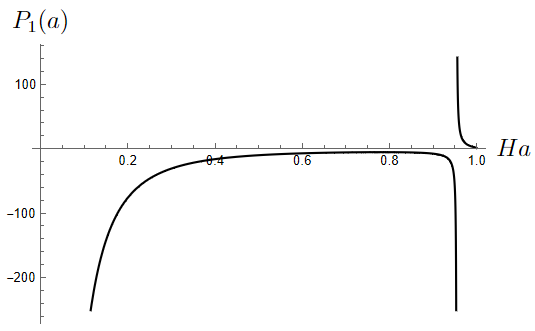}
    \caption{}
    \label{fig-5b}
  \end{subfigure}
  \caption{A plot of $P_{n}(a)$ in the entire Euclidean region for the modes $n=1,2,3,4,5$ and $\mu=3/2$.} 
  \label{fig5}
\end{figure}

\begin{figure}[h!]
		\includegraphics[scale=0.5]{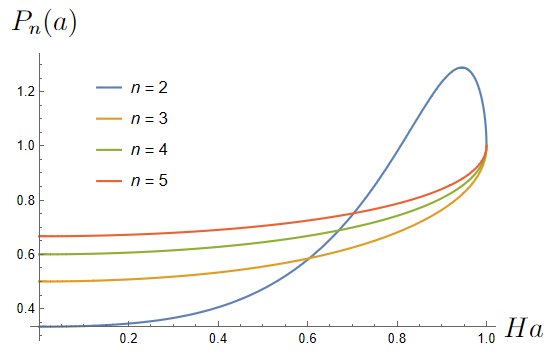}
		\caption{A plot of $P_{n}(a)$ in the entire Euclidean region for the modes $n={2,3,4,5}$ and $\mu=1.75$. }
		\label{fig6}
 
\end{figure}

To illustrate the characteristic behavior of the ratio $P_{n}(a)$, we plotted it in Fig.\ref{fig-5a} in the Euclidean region $Ha<1$ for the leading inhomogeneous modes with $\mu=3/2$. It is always less than $1$, apart from the turning point $Ha=1$, where $P_n(H^{-1})=1$. By sampling different values of $\mu$, we found that his behavior holds in the entire range of $\mu\in(0,2)$ for all modes, with the following two exceptions.  One exception is the homogeneous mode $n=1$ which exhibits a segment adjacent to the turning point where $P_{1}(a)>1$ for all $\mu>0$ (see Fig.\ref{fig-5b}).  The other is the dipole mode $n=2$ which exhibits similar behavior for $\mu\gtrsim1.6$ (see Fig.\ref{fig6}).

For modes violating the condition (\ref{P<1}), we cannot trust their behavior on the growing branch in the classically forbidden region.  For the homogeneous mode this means that violation of regularity by $R_1^-(a)$ may be spurious in this case.  Note also that even if $R_1^-(a) <0$ in some range of $a<H^{-1}$, this would not necessarily be a problem.  The homogeneous mode on the growing branch behaves as the Hartle-Hawking wave function which has a minimum at the maximum of $V(\phi)$.  This would not be problematic, as long as the growing branch remains subdominant.

For the dipole mode, violation of (\ref{P<1}) implies that we could not have used regularity of $R_2^-(a)$ to select the mode function $\nu_2(a)$ for $\mu\gtrsim1.6$.  This suggests that in this range of $\mu$ a different dipole mode function could be selected which would satisfy the regularity condition. We have verified that this is indeed the case for mode functions (\ref{modes}) with a range of values of $B_2/A_2$.  We note however that a specific value of $B_2/A_2$ is not selected in this case. The significance of the anomalous behavior of the dipole mode is not clear to us.  This issue may require further study.

\section{Field variance}

In this section we will calculate the individual variances for each mode and the total variance of the field by determining the values of $Re(R_{n}(a))$. We will first consider the inhomogeneous modes $n>1$. An explicit calculation yields:
\beq
Re(R_{n}(a))=-\frac{Ha^{3}}{h_{1}+h_{2}+h_{3}}
\label{Rh}
\eeq
where the functions $h_{i}$, $i=1,2,3$ are given respectively by:
\beq
h_{1}\sim (Ha)^{3+2\gamma}\frac{\Gamma\left(-1-\gamma+n\right)}{\Gamma\left(2+\gamma+n\right)},
\eeq
\beq
h_{2}\sim1,
\eeq
\beq
h_{3}\sim (Ha)^{-(3+2\gamma)}\frac{\Gamma\left(2+\gamma+n\right)}{\Gamma\left(-1-\gamma+n\right)},
\eeq
where we factored out $\mathcal{O}(1)$ terms independent of $n$ and $a$.

For $n\gg 1$ we can use the asymptotics of Gamma functions to find
\beq
h_1\sim \left(\frac{Ha}{n}\right)^{3+2\gamma},
\label{h1}
\eeq
\beq
h_3\sim \left(\frac{n}{Ha}\right)^{3+2\gamma}.
\eeq
(Note that the range of interest for $\gamma$ is $0<\gamma<1$.)
It follows that for $n\ll n_*\sim Ha$ we have $h_1\gg h_2, h_3$; hence we can drop $h_2$ and $h_3$ in Eq.(\ref{Rh}).  Then using Eqs.(\ref{vari}),(\ref{h1}) we obtain
\beq
\sigma_n^2\sim H^2 n^{-3-2\gamma}(Ha)^{2\gamma}.
\label{sigman}
\eeq
Since $\gamma>0$, successive terms in the sum of Eq.(\ref{sigmatot}) decrease faster than $n^{-1}$, and thus the sum is not sensitive to the cutoff $n_*$.  For $n\sim n_*$, we have $h_1\sim h_2\sim h_3$, but this does not change the estimate (\ref{sigman}) by order of magnitude.  In any case, the estimate of $\sigma_n$ at large $n$ is unimportant, since the large-$n$ terms give a negligible contribution to the sum.

Including now $O(1)$ factors, we find 
\beq
\sigma_{n}^{2}=\frac{\pi\Gamma\left(\gamma+2+n\right)}{4^{\gamma+1}|\Gamma\left(-\gamma-1+n\right)|\left[\Gamma\left(\gamma+\frac{3}{2}\right)\right]^{2}}\frac{(Ha)^{-2\gamma}}{H^{2}}
\eeq
where the absolute value is necessary to include the homogeneous mode $n=1$.  To a good approximation we can extend the summation in Eq.(\ref{sigmatot}) to $n\to\infty$.  This gives
\beq
\sigma^{2}\approx CH^{2}(Ha)^{2\gamma},
\label{varianceC}
\eeq
where
\beq
C=\frac{2^{1+2\gamma}}{\sin(\gamma\pi)}\frac{\left[\Gamma\left(\gamma+\frac{3}{2}\right)\right]^{2}}{\Gamma\left(\gamma+1\right)\Gamma\left(\gamma+3\right)}
\eeq

\end{appendices}

\end{document}